\documentclass[12pt]{article}
\def\demi{{\textstyle {1\over2}}}

\let\w=\wedge
\let\o=\omega
\let\O=\Omega

\def\L{ {\cal L}}
\def\vo{{\mathcal V}}

\def\be{\begin{equation}}
\def\ee{\end{equation}}

\begin{document}
\bibliographystyle{perso}

\begin{titlepage}
\null \vskip -0.6cm
\hfill PAR--LPTHE 02--06

\hfill hep-th/0207020

\vskip 1.4truecm
\begin{center}
\obeylines

        {\Large Eight-Dimensional Topological Gravity
        and its Correspondence with Supergravity}
\vskip 6mm
Laurent Baulieu$^a$, M. Bellon $^a$ and  Alessandro Tanzini$^{a,b}$
{\em $^a$ Laboratoire de Physique Th\'eorique et Hautes Energies,
 Universit\'es Pierre et Marie Curie, Paris~VI
et Denis~Diderot,~Paris~VII}
{\em $^b$ Istituto Nazionale di Fisica Nucleare, Roma}

\end{center}

\vskip 13mm

\noindent{\bf Abstract}: A topological theory for euclidean gravity in eight
dimensions is built by enforcing octonionic self-duality conditions on the
spin connection.
The eight-dimensional manifold must be of a special type, with $G_2\subset 
Spin(7) \subset SO(8) $ holonomy.
The resulting theory is related to a twisted version of $N=1$, $D=8$
supergravity. 
The situation is comparable to that of the topological Yang--Mills theory
in eight dimensions,
for which the $SO(8)$ invariance is broken down to $Spin(7)$, but is recovered
after untwisting the topological theory. 
\vfill

\begin{center}

\hrule \medskip
\obeylines
Postal address: %
Laboratoire de Physique Th\'eorique et des Hautes Energies,
 Unit\'e Mixte de Recherche CNRS 7589,
 Universit\'e Pierre et Marie Curie, bo\^\i te postale 126.
4, place Jussieu, F--75252 PARIS Cedex 05

\end{center}
\end{titlepage}
\def\w{\wedge}
\def\o{\omega}
\def\t{\tilde}
\def\TQFT{ Topological Field Theory}
\section{Introduction}

We have shown in~\cite{BaTa} that topological gravity in four dimensions can
be identified with the $N=2$, $D=4$
supergravity, for manifolds with special holonomy.  
Our inspiration was that superstring theories can be obtained, at least in
a formal way, as suitable
anomaly free untwisting of topological sigma-models~\cite{green}, so that 
supergravities, which arise as low energy limits of superstrings, should
be linked to topological gravities.

Here we consider  the case of eight-dimensional gravity, with the ultimate
goal of showing that $D=11$ supergravity,
which   determines all known supergravities in lower dimensions, can   be
viewed as a  topological theory. 
We indicate in this note  some very encouraging results in this direction.  We
show how the Einstein action plus the Rarita--Schwinger term can be obtained
in eight dimensions by imposing in a BRST invariant way the octonionic
gravitational instanton equation, generalizing the $N=2$, $D=4$
supergravity case worked out in~\cite{BaTa,Fre}.  As was shown in~\cite{bakasi},
going from four dimensions to eight dimensions  amounts to changing the
quaternionic structure coefficients
into octonionic ones in the self-duality equations that are used to express
the topological field theory. 
From the beginning, this imposes that one uses a manifold with   special
holonomy group, 
$H\subset SO(8)$. Actually,  we will reach the conclusion   that the
topological field theory
must be considered on a manifold with  $G_2 \subset Spin(7) \subset SO(8)$
holonomy. 

As we will discuss in sect.~4, the topological model that we consider
is related to a twisted version of $N=1$, $D=8$ supergravity.  It thus
can be viewed as an effective supergravity of heterotic string theory. This is
in line with the results of~\cite{Billo}, where the twist of the four
dimensional effective theories of heterotic superstrings has been
discussed.  We leave to another paper a delicate point, the fact that
the presence of unpaired fields in the topological model should require
the introduction of a new sector involving a two-form gauge field, 
completing the identification with supergravity.  This question is
under consideration but its technical complications deserve a separate
publication.

\section{The octonionic self duality equation and the Einstein action in eight
dimensions}

As in four dimensions, we would like to build a topological field theory by
enforcing self-duality conditions on the spin connection $\o$, taken as a
functional of the vielbein $e$~\cite{BaTa}. 
In eight dimensions, a natural choice is the octonionic gravitational
instanton equation~\cite{floratos}, given by
\begin{equation}\label{octo}
\o^{ab}-\demi \O^{abcd} \o^{cd}=0,
\end{equation}
where $\O$ is a completely antisymmetric four-tensor invariant under a
$Spin(7)$ subgroup of $SO(8)$.  Indeed, the four-tensor $\O$ induces a
$Spin(7)$ invariant decomposition of the adjoint representation of $SO(8)$,
${\bf 28=7} \oplus {\bf 21}$~\cite{bakasi}. Eq.~(\ref{octo}) corresponds to
setting to zero the components $\o^{ab^-}$ of the spin connection in the seven
dimensional subspace; it then counts for $56=7\times 8$ conditions, which
exactly match the number of degrees of freedom
contained in the vielbein $e_\mu^a$ modulo reparametrizations.

A remarkable fact is that, as happens in the four-dimensional
case~\cite{BaTa},
the eight-dimensional Einstein Lagrangian can be expressed, up to a pure
derivative, as a quadratic form in the anti-self-dual part of the spin
connection:
\begin{eqnarray}\label{selfdual}
\L_{EH}  &=&{1\over 2 }  R ^{ab} \vo_{ab} = 2 \;\o^{ac^-} \o^{bc^-} \vo_{ab} +
2\;d(\o^{ab^-} \vo_{ab}),
\end{eqnarray}
where $R$ is the Lorentz curvature and 
\begin{equation}
\vo_{ab}= {1\over 6!} \epsilon_{abcdfghj} e^ce^de^fe^ge^he^j.
\end{equation}
Notice that each of the two terms in the right hand side of eq.~(\ref{selfdual}) 
explicitly displays only $Spin(7)$ gauge invariance.  Thus, this expression of
the Einstein action is globally well defined only on manifolds of $Spin(7)$ holonomy.
From eq.~(\ref{selfdual}) follows that
any solution of eq.~(\ref{octo}) is a stationary point of the Einstein
action and a solution of the Einstein equations.   This suggests that the
topological field theory that we shall build using eq.~(\ref{octo}) as gauge
fixing contains ordinary gravitation.

The proof of eq.~(\ref{selfdual}) starts from the Bianchi identity,
$DT^a=R^{ab}e^b=0$, which implies $\O R^{ab} e^b e^a=0$.
This can be rewritten as 
\begin{equation}\label{bianchi}
\O_{abcd}R^{cd}\vo^{ab}=0,
\end{equation}
using the fact  that
$\O$ is a self dual 4-form. We then decompose $R^{ab}$ under $Spin(7)$,
writing 
\begin{equation}\label{deco}
R^{ab}= R^{ab^+}+ R^{ab^-}, \quad \demi \O_{abcd} R^{ab}= R^{cd^+}-3 R^{cd^-}.
\end{equation}
Eq.~(\ref{deco}) and the Bianchi identity expressed in eq.~(\ref{bianchi})
allows for the elimination of $R^{ab^+}$ in the Einstein Lagrangian:
\begin{equation}
{1\over2}R^{ab} \vo_{ab}
  =2R^{ab^-} \vo_{ab}
\end{equation}
Using that $R^{ab^-}=d\o^{ab^-}+\demi[\o,\o^-]^{ab}
- \o^{ac^-}\o^{cb^-}$ and the zero-torsion condition, one finally gets the very
interesting  identity of eq.~(\ref{selfdual}).

\section{The fields of topological gravity }   
We now proceed to the construction of the topological field theory
based on the gauge-fixing condition~(\ref{octo}).
The set of fields of the TQFT  is the eight-dimensional generalization of that
used in~\cite{BaTa}:
\begin{eqnarray}\label{viel}
&&\matrix
{   &    &  e^a_\mu  &   &  \cr
    &  \Psi^a_\mu  &       &  \bar\Psi^{ab^-}_\mu & \cr
    \Phi^a  &    &       L^{ab^-},b^{ab^-}_\mu & & \bar\Phi^a  \cr
    &  \eta^{ab^-} &         &     \bar \eta^{a } & \cr }
\\
\noalign{\bigskip}
\label{spin}
&&\quad\matrix{
&    &\o^{ab}_\mu  &   &   \cr
&   \tilde\Psi^{ab}_\mu  &    & \bar{\tilde\Psi}^{ab}_\mu \cr
 {\tilde \Phi}^{ab } &    & {\tilde L}^{ab },{\tilde b}^{ab}_\mu &   &
\bar{\tilde\Phi}^{ab} \cr
&    {\tilde\eta}^{ab }  &   &       \bar {\tilde\eta}^{{ab} }& \cr }
\\
\noalign{\bigskip}
\label{gauge}
&&\quad\quad\matrix
{ &    &  A_\mu  &   & \cr
&  \bar \Psi^{}_\mu  &  &   {\chi}^{ab^-} \cr
\Phi &      &  \beta^{ab^-} & & { \bar \Phi}\cr
   &     &   &   { \eta}^{ }\cr }
\\
\noalign{\bigskip}
\label{ghost}
\matrix
{ \xi^ \mu   &     &    \bar \xi^ \mu \cr
     &    b^\mu }
&&\quad\quad\quad\quad\quad
\matrix
{ \O^ {ab}  &   &        \bar \O^ {ab} \cr
     &  b^{ab}  & }
\quad\quad
\quad \quad\quad\matrix
{     c   &   &        \bar c \cr
    &  b   & }
\end{eqnarray}
In the above tables, ghost numbers increase to the left and the BRST
transformation acts in the south-west direction.

The Lorentz$\times$diffeomorphism symmetry determines a Lie group and
its associated topological BRST symmetry can be constructed as:
\begin{eqnarray}
se^a_\mu &=& \L_\xi e^a_\mu -\O^{ab} e^b_\mu +\Psi^a_\mu \nonumber \\
s\o^{ab}_\mu &=& \L_\xi \o^{ab}_\mu +D_\mu\O^{ab}   +\t \Psi^{ab}_\mu
\nonumber\\
\noalign{\medskip}
s\Psi^a_\mu &=& \L_\xi  \Psi^a_\mu -\O^{ab} \Psi^b_\mu-\L_\Phi e^a_\mu
+\t \Phi^{ab}e^b_\mu
\nonumber\\
s \t \Psi ^{ab}_\mu &=& \L_\xi  \t\Psi ^{ab}_\mu - [\O ,\t \Psi_\mu]^{ab}  
+ D_\mu \t \Phi^{ab}
-\L _\Phi \o^{ab}_\mu 
\nonumber\\
\noalign{\medskip}
s  \Phi ^{a}  &=& \L_\xi   \Phi ^{a } -\O ^{ac} \Phi^{c}  
\nonumber\\
s \t \Phi ^{ab}  &=& \L_\xi  \t\Phi ^{ab} - [\O , \t \Phi]^{ab}  
\nonumber\\
\noalign{\medskip}
s\xi^\mu  &=& \Phi^\mu+\xi^\nu \partial _\nu \xi^\mu
\nonumber\\
s \O ^{ab} &=& \L_\xi  \O ^{ab} 
-\O ^{ac}\O^{cb}  + \t \Phi^{ab} \ \ .
\label{brs} 
\end{eqnarray}
Moreover, we have that $\Phi^a=e^a_\mu \Phi^\mu$, and thus,
$s  \Phi ^{\mu} = \L_\xi   \Phi ^{\mu } $. 
The other fields in eq.~(\ref{viel}), (\ref{spin}) and~(\ref{ghost})
are doublets of antighosts and Lagrange multipliers.
Their BRST transformations are given by
\def\s{\hat s}
\def\bg{\bar  g}
\begin{eqnarray}
\s \bg = \lambda \quad  \s \lambda =\L_\Phi \bg + \delta_{\t \Phi} \bg
[\t \Phi,\bg] \ \ ,
\end{eqnarray}
with  $s X= \s X +\L_\xi X+\delta_{\O} X$ and $\delta_{\O}$
a Lorentz transformation with parameter $\O$.

Finally, the BRST transformations for the fields of the 
$U(1)$ gauge sector are 
\begin{eqnarray} 
sA_\mu&=&    \L_\xi A_\mu+\partial_\mu  c +\bar \Psi_\mu
\cr s\bar \Psi_\mu&=& \L_\xi \bar \Psi_\mu -\partial_\mu  \Phi
-\L_{\Phi}A_\mu
\cr s \Phi &=&    \L_\xi  \Phi - \L_{\Phi} c
\cr sc  &=&    \L_\xi c+ \Phi \ \ .
\end{eqnarray}
The fields in eqs.~(\ref{viel}--\ref{ghost}) 
have a suitable decomposition
under $Spin(7)$, according to the choice of the gauge-fixing
condition~(\ref{octo}). 

The fields $(\bar\Psi_\mu^{ab^-}, b_\mu^{ab^-})$ in~(\ref{viel})
are respectively the antighost and Lagrange multiplier associated 
to the gauge-fixing condition~(\ref{octo}), and thus belong to the seven
dimensional anti-self-dual representation of $Spin(7)$.
Moreover, the presence of the $U(1)$ topological multiplet implies
a further decomposition of the twenty-one-dimensional self-dual 
representation of $Spin(7)$ under a $G_2$ subgroup,
${\bf 21=7} \oplus {\bf 14}$. The fields $\chi^{ab^+}$ and 
$\beta^{ab^+}$ in~(\ref{gauge}) belong to the seven-dimensional
subspace of this $G_2$ decomposition.

In the following section we will describe the topological theory
that can be built from the fields in eqs.~(\ref{viel}--\ref{ghost})
and its relationship with a suitable twisted version of 
$N=1$, $D=8$ supergravity.

\section{Topological action and its relationship with twisted supergravity}

In order to implement eq.~(\ref{selfdual}), we consider the following action,
for a manifold with $Spin(7)$ holonomy:
\begin{eqnarray}\label{act}
I &=&\int_{{\cal M}_8}  s \Big(\bar \Psi ^{ab^-}
(\o^{ac^-}+b^{ac^-})\vo_{bc}
+ F \Psi^a e^a\Big)
\end{eqnarray}
The action eq.~(\ref{act}) defines 
a cohomological theory,
whose BRST symmetry is associated to a
$Spin(7)\times$diffeomorphism$\times$gauge invariance.  
By expanding the first term of eq.~(\ref{act}) one gets the bosonic action 
\begin{equation}\label{b-act}
I_1=\int_{{\cal}_ {{\cal M}_8}} 
 b ^{ab^-}b^{ac^-} \vo_{bc}
+b ^{ab^-}\o^{ac^-} \vo_{bc} \ \ ,
\end{equation}
which reproduces the Einstein
action after elimination of the field $b$, 
according to the identity~(\ref{selfdual}).
A natural question now arises, whether the topological model
we are considering can be identified with a suitable twisted version
of eight-dimensional supergravity. This turns out to be actually the case.
Notice in fact that on a manifold with $Spin(7)$ holonomy there exist
a covariantly constant spinor (of norm one) $\eta$, which can be used to 
redefine 
the gravitino $\bar\Lambda=(\lambda, \bar\lambda)$ of $N=1$, $D=8$ 
supergravity as
\begin{eqnarray}\label{twist}
\lambda &=& \Psi^a \gamma_a \eta \ \ , \nonumber\\ 
\bar\lambda &=& \bar\Psi \eta + \bar\Psi^{ab^-}\gamma_{ab}\eta \ \ ,
\end{eqnarray}
where $(\lambda, \bar\lambda)$ are Weyl spinors
of opposite chiralities and the eight-dimensional gamma matrices 
$\gamma_a$ acts on spinors of definite chirality.
Moreover, we have ~\cite{bakasi}
\begin{equation}
\O_{abcd}=\eta^T \gamma_{abcd}\eta .
\label{omega}
\end{equation}
From eq.~(\ref{twist}) we see that the gravitino is then mapped
to the fields $(\Psi^a, \bar\Psi, \bar\Psi^{ab^-})$ of the topological model.
This suggest to compare the topological action~(\ref{act}) with
a twisted version of $N=1$, $D=8$ supergravity.
In the spirit of topological field theory, we can restrict our attention
on the kinetic terms for the fields, which simplifies considerably
the comparison. In fact, the basic requirement 
on the gauge-fixing conditions is that they must give a good definition
for the propagators of the fields. Interaction terms can be then always
added as BRST exact terms in order to get agreement with the complete 
twisted supergravity action.
The kinetic terms for the fields $(\Psi^a, \bar\Psi, \bar\Psi^{ab^-})$
in the topological action eq.~(\ref{act}) are given by 
\begin{eqnarray}\label{f-act}
I_{2}&=&\int_{{\cal}_ {{\cal M}_8}}\Big[ \sqrt{g}\  d^8 x \ \ 
\Big(\bar \Psi ^{ac^-}_a (\partial_{[b}\Psi_{c]}^b
+{1\over 2} \O_{cbfg }\partial _b\Psi ^g_f )
-{1\over 2} \Psi ^{ac^-}_b \partial_{[a}\Psi_{c]}^b \nonumber \\ 
&& - {1\over 2} \bar \Psi ^{ac^-}_b
\O_{cbfg }(
\partial _{[a}\Psi ^g_{f ]} -
\partial _{f}\Psi ^a_{g} )\Big)
+ \  \O \ \bar\Psi d\Psi^a e^a \Big]
\end{eqnarray}
After some 
algebra on the gamma matrices and using 
eq.~(\ref{omega}) and the self-duality
of the four-form $\O$, 
one can verify that the action eq.~(\ref{f-act})
corresponds to the quadratic part of the ordinary
Rarita-Schwinger action
\begin{equation}\label{rs} 
I_{RS}=\int  \sqrt{g}\  d^8 x \ \ 
\bar\Lambda_a \Gamma^{abc}\partial_b \Lambda_c \ \ ,
\end{equation}
written in terms of the twisted variables eq.~(\ref{twist}).

Finally, we add to the action eq.~(\ref{act})
the term
\begin{equation}\label{gaugeact}
I_{gauge}= \int_{{\cal M}_8} \ s \Big[ \O \ \chi^{+}(\beta^{+}+ F)\Big]
\end{equation}
which,  upon integration on the auxiliary field $\beta^+$,
gives the right kinetic term for the graviphoton field
$A$~\cite{bakasi}. Notice that the presence in eq.~(\ref{gaugeact})
of the fields $(\chi^+, \beta^+)$ breaks the $Spin(7)$
symmetry down to $G_2$.

Summarizing, we have seen that the topological model defined 
by the action eq.~(\ref{act}) and eq.~(\ref{gaugeact}) corresponds to the twisted
version of a gravitational model containing a graviton,
a gravitino and a graviphoton. 
In order to make a full comparison
with $N=1$, $D=8$ supergravity, 
we should still introduce in the topological field theory 
a sector describing a two-form
field and the corresponding gauge field
entering the definition of the curvature three-form~\cite{sase}.
We deserve the analysis of this sector for further work.

Here we only remark that the gauge-fixing of the $Spin(7)$
symmetry of our model already gives
indications on the presence of a further topological multiplet
in order to have a well defined functional integral for all the fields
introduced in sect.3.
In fact, in defining the topological action, one must use all fields 
that occur in the topological quartets in eq.~(\ref{viel}-\ref{ghost}). 
Let us describe in detail how this is done.

First of all, the topological freedom in $\o$ is used  to constrain the
torsion
$T=de+\o e$ and determine  $\o$ as a functional of the vielbein $e$.
This allows one to eliminate the fields  
$\bar{\tilde \Psi}^{ab}$ and   $ {\tilde \Psi}^{ab}$.

Then, since the octonionic self-duality equation~(\ref{octo}) is only 
$Spin(7)\subset SO(8)$ gauge invariant, one uses 
seven freedoms contained in the Lorentz antighost 
$\bar{\tilde \Phi}$ to enforce that $\Omega^{ab^-}=0$. 

As discussed before, the graviphoton Maxwell term $|F|^2$
is generated by the action in eq.~(\ref{gaugeact}). To eliminate
the fermionic partner $\chi^+$ of  
$\beta^{+}$, one must then use 
seven other freedoms contained in $\bar{\tilde\Phi}$.  
As evocated earlier, this imposes that one goes to a further decomposition
under a $G_2$ subgroup of $Spin(7)$, which gives
$\bar{\tilde
\Phi} ^{ab}=  \bar{\tilde
\Phi}_7 ^{ab^-} 
+\bar{\tilde
\Phi}_7^{ab^+}+\bar{\tilde
\Phi}_{14}^{ab^+}$.
We then add to our action the following $s$-exact term:
\begin{equation}\label{deb} 
 \int_{{\cal}_ {{\cal M}_8}}\  s \Big[ \sqrt{g} \
\Big(\bar{\tilde \Phi}_7^{ab^-} \O^{ab^-} 
+
\bar{\tilde \Phi}_7^{ab^+} ({\chi}^{ab^+} - \O_7^{ab^+})
 \Big)\Big]
\end{equation}
The gauge symmetry which arise for the fields $(\Psi^a, \bar \Psi,
\bar \Psi^{ab^-})$, is fixed by using the antighost 
$(\bar \Phi^a, \tilde L^{ab^-},\bar \Phi)$ and their fermionic
Lagrange multipliers exactly as in~\cite{BaTa}, with the $s$-exact action:
\be\label{gfghost}
I_{ghosts}=
\int d^8x \
s\Big[\sqrt{g}(\bar \Phi ^a   D_\mu   \Psi ^{a }_\mu+
{\t L}^{ab^-}  D_\mu   \bar \Psi ^{ab^-}_\mu
+\bar \Phi \partial_\mu   \bar \Psi_\mu)\Big]
\ee
This term, after untwisting, provides the 
gauge-fixing for the longitudinal parts of all components of the gravitino. 
This indicates that the fields
$(\bar \Phi^a,  \tilde L^{ab^-},\bar\Phi)$ can be identified as 
the commuting Faddeev-Popov 
antighosts of
$N=1,D=8$ local supersymmetry, while the fields $(\bar \eta^a, {\tilde
\eta}^{ab^-}, \eta)$ can be identified as 
the corresponding anti-commuting Nielsen-Kallosh ghosts.

After all, we see that the field
$\bar{\tilde \Phi}_{14}^{ab}$ 
in the topological 
multiplet of the spin connection eq.~(\ref{spin}) is still not exploited. 
This is the signal that another topological
multiplet can be introduced, whose gauge fixing conditions can be imposed using
these remaining degrees of freedom.
The mechanism here is analogous to the introduction
of the graviphoton $A$ in order to complete the topological
gauge-fixing of the vielbein $e$ and of the spin connection $\o$
~\cite{BaTa} . 
In the present case the extra topological
multiplet should be naturally identified with that of a two-form gauge field. 

From the geometrical point of view, the presence of a two-form
should induce a non trivial torsion, and 
the breaking to a $G_2$ holonomy group could be expected
since this gives a natural way to define a torsion tensor
$S^{ijk}\sim c^{ijk}$ proportional to the octonionic 
structure constants $c^{ijk}=\O^{ijk8}$, with $i,j,k=1,\ldots,7$.
We reserve to further work this construction.

\section{Conclusion}

In this paper we have shown  that the determination of a gravitational
topological field theory can be generalized from the case of four dimensions
to that of eight dimensions.  It is interesting to remark how the BRST
transformations in our topological model generates local supersymmetry.  
In
fact, this model appears related to a twisted version of
$N=1$, $D=8$ supergravity. Although the situation is technically more
involved than in four dimensions, the twist operation is of a
conceptually simpler nature in eight dimensions. This is due to the
existence of triality, which allows one to simply identify spinor and
vector indices. The twist operation appears as a change of variables that 
can be defined on  manifolds with holonomy group
contained in $Spin(7)$.  Such manifolds with $G_2$ and $Spin(7)$ holonomy
have recently attracted a renewed attention in the context of
$M$-theory compactifications~\cite{gibb}.  An interesting result of our
analysis is that on these manifolds the Einstein action can be written
as a quadratic form in the anti-self-dual part of the spin connection,
eq.~(\ref{selfdual}).  The relevant r\^ole, which is played
in this context by the self-duality conditions on the spin connection,
has been underlined also in~\cite{bilal, sp7}.

The observables of our topological model could be defined from the solution
of descent equations starting from the eight form $\O R^{ab} R^{ab} $.
It remains to identify such observables with known or unknown topological
invariants of eight dimensional manifolds of special holonomy.

We find this result as a very encouraging step toward our goal which is to
show that  $N=1$, $D=8$ supergravity can be identified as an untwisted
topological field theory. It must be noted that the dimensional reduction of
our model in lower dimensions indicate 
that the low energy effective action of $N=2$, $D=4$ supergravity
could be analysed \`a la Seiberg and Witten, in a way that
parallels the treatment of the Yang--Mills theory of~\cite{bakasi}.



\end{document}